\documentstyle[12pt]{article}

\newcommand{\resection}[1]{\setcounter{equation}{0}\section{#1}}

\def\s {\sigma}
\def\e {\epsilon}
\def\be{\begin{equation}}
\def\ee{\end{equation}}
\def\ba{\begin{array}}
\def\ea{\end{array}}
\def\EQ{\begin{equation}}
\def\EN{\end{equation}}
\def\bea{\begin{eqnarray}}
\def\eea{\end{eqnarray}}
\def\to{\rightarrow}
\def\goto{\longrightarrow}
\def\sa{\hspace{0.1in}}
\def\hs{\hspace{0.1in}}
\def\sb{\hspace{0.2in}}

\def\C{{\cal C}}
\def\K{{\cal K}}
\def\M{{\cal M}}
\def\O{{\cal O}}
\def\P{{\cal P}}
\def\D{\Delta}
\begin{document}
\oddsidemargin 5mm
\setcounter{page}{0}
\renewcommand{\thefootnote}{\fnsymbol{footnote}}
\newpage
\setcounter{page}{0}
\begin{titlepage}
\begin{flushright}
DAMTP-HEP-94/15
\end{flushright}
\vspace{0.5cm}
\begin{center}
{\large {\bf Form-Factor Bootstrap and the\\
 Operator Content of Perturbed Minimal Models }} \\
\vspace{1.5cm}
{\bf
A. Koubek\footnote{E-mail: a.koubek@amtp.cam.ac.uk} }\\
\vspace{0.8cm}
{\em Department of Applied Mathematics and Theoretical Physics,\\
Silver Street,\\
CB3 9EW Cambridge, \\UK} \\
\end{center}
\vspace{6mm}
\begin{abstract}
 The form-factor bootstrap approach is applied to the perturbed
minimal models $M_{2,2n+3}$ in the direction of the primary field
$\phi_{1,3}$. These theories are integrable and contain $n$ massive
scalar particles, whose $S$--matrix is purely elastic.  The
form-factor equations do not refer to a specific operator. We use this
fact to classify the operator content of these models. We show that
the perturbed models contain the same number of primary fields as the
conformal ones. Explicit solutions are constructed and conjectured to
correspond to the off-critical primary fields $\phi_{1,k}$.
\end{abstract}
\vspace{5mm}
\begin{center}
May 1994
\end{center}
\vspace{10mm}
\begin{center}
revised version
\end{center}
\end{titlepage}
\newpage

\resection{Introduction}

The identification of local operators in a given model and the
computation of their multipoint correlators is one of the most
important problems in Quantum Field Theory (QFT). For two-dimensional
critical QFT, a successful approach is given by the conformal
bootstrap analysis \cite{Polyakov,BPZ,FQS,Cardy,Felder,ISZ} . The
operator content of such theories consists of a set of conformal
families which form irreducible representations of the conformal
Virasoro algebra.  The representatives of these families, organized in
a tower of {\em descendant operators} of increasing anomalous
dimensions and spin, are identified by their {\em primary operators}.
The number of conformal families is finite for the so-called minimal
models. Using the representation theory of the underlying Virasoro (or
higher dimensional) algebra, it turns out that the correlation
functions of conformal operators in the minimal models can be written
as solutions of linear differential equations \cite{BPZ,Dotsenko}.

Concerning the massive integrable QFT, the present understanding is
incomplete, though much progress has been achieved recently.  The
on--shell behaviour of these theories is characterized by the
factorizable $S$-matrix of the asymptotic states \cite{ZZ,Zam,Musrep}.
Several authors have pursued the idea that these QFT (or analogous
off-critical lattice models) can be characterized by infinite
dimensional algebras, which generalize the Virasoro algebra of the
critical points and  lead to the computation of
correlation functions by exploiting this algebraic structure
\cite{LeClair,Sotkov,lyk}. An alternative non--perturbative
method in which the algebraic structure is less apparent is known as
{\em form factor bootstrap approach}
\cite{Smirnov1,smirnov13,YLZam,CMform,FMS}. Using general properties
of unitarity, analyticity and locality, this approach leads to a
system of functional and recursive equations for the matrix elements
of local operators between asymptotic states which allow their
explicit determination. Using a parametrization of the external
momenta in terms of the rapidity $\beta$
\[
p_i^0\,=\,m_i\cosh\beta_i\,\,\, , \,\,\,
p_i^1\,=\,m_i\sinh\beta_i\,\,\, ,
\]
and assuming CPT invariance, matrix elements of a generic local
operator ${\cal O}$ can be cast in the following form
\EQ
F_{\e_1,\e_2,\dots,\e_n}^{\cal O} (\beta_1,\beta_2,\ldots,\beta_n)
\,=\,
\langle 0\mid{\cal O}(0,0)\mid Z_{\e_1}( \beta_1),
Z_{\e_2}(\beta_2),\ldots,Z_{\e_n}(\beta_n)\rangle_{in}\sb ,
\label{FoF}
\EN
called the {\em form factors}. The important point is that the
functional equations satisfied by the form factors do {\em not} refer
to a specific operator. This physical arbitrariness is determined by
the mathematical property of the existence of different solutions to
the same set of functional equations. This fact opens the possibility
to classify the operator content of the QFT under investigation by
identifying the solution space of the form factor equations.  In fact
the space of the solutions is isomorphic to the space of local
operators entering the QFT.

Besides the classification of the operator content the form--factor
bootstrap approach gives a means of calculating correlation functions
in the QFT under consideration.  To see this, let us consider the
two-point correlators\footnote{Similar consideration can be easily
extended to multi-point correlation functions.}

\[
G_i(x)\,=\,<\O_i(x)\,\O_i(0)> \sb ,
\]
 of hermitian operators. It can be expressed as an infinite series
over multi-particle intermediate states
\be
\langle \O_i(x)\,\O_i(0)\rangle\,=\label{correlation}\ee
$$ \sum_{n=0}^{\infty}
\int \frac{d\beta_1\ldots d\beta_n}{n! (2\pi)^n}
<0|\O_i(x)|Z_{\e_1}(\beta_1),\ldots,Z_{\e_n}(\beta_n)>_{\rm in}{}_{\rm
in} <Z_{\e_1}(\beta_1),\ldots,Z_{\e_n}(\beta_n)|\O_i(0)|0> = $$
$$\sum_{n=0}^{\infty}
\int \frac{d\beta_1\ldots d\beta_n}{n! (2\pi)^n}
\vert F_{\e_1,\e_2,\dots,\e_n}^{\cal O}
(\beta_1,\beta_2,\ldots,\beta_n) \vert^2 e^{-mr \sum_i \cosh \beta_i}
$$ where $r$ denotes the radial distance in the euclidean space, {\em
i.e.} $r=\sqrt{x_0^2 + x_1^2}$. It is evident from this expression
that the form factor expansion plays the role of a {\em partial wave}
decomposition of the correlators and therefore their knowledge
provides the knowledge of the correlation--function in terms of a
convergent series expansion.

We will examine the form--factor bootstrap approach for the perturbed
minimal models $\M_{2,2p+3}+\Phi_{1,3}$\footnote{The form--factors for
some special operators of these models have been calculated by
different methods in \cite{Smirnov1,smirnov13} and the first model of this
series, the perturbed Yang--Lee model has been analyzed in
\cite{YLZam}.}. The Kac-table of these non--unitary conformal theories
extends along a row, and the model $\M_{2,2p+3}$ contains exactly $p$
primary fields. The perturbations of these models along the
$\Phi_{1,3}$ directions are integrable \cite{Zam} and can be described
as restrictions of the Sine-Gordon model \cite{smirnov13,EguchiYang}.

We choose this series of models because their $S$--matrix contains
only scalar particles and they thus have a simple fusion algebra of
massive particles.  The on--shell $S$--matrices are given by
\cite{Chicago}
\be
S_{ab}=f_{\mid a-b\mid\frac\alpha 2}(\beta) f_{(a+b)\frac \alpha
2}(\beta)
\prod_{k=1}^{min(a,b)-1} (f_{(\mid a-b\mid +2k)\frac\alpha 2}(\beta))^2
\sb , \label{sss}
\ee
where $\alpha=\frac {2\pi}{2p+1}$, and $a,b=1,2,\ldots p$ labels
the particles of mass $m_a=\sin\left (a \frac{\alpha}{2}\right )$.
The functions $f$ are given by
\be f_\alpha(\beta) \equiv \frac{\tanh \frac 12(\beta+i\alpha)}{\tanh \frac 12
(\beta-i\alpha)} \sb .\ee The particles obey the bootstrap
fusion algebra
\be \ba{l}
a_i \times a_j \to a_{i+j}\sb\sb {\rm or} \\ a_i \times a_j \to
a_{2p+1-i-j}\sb\sb {\rm and} \\ a_i \times a_j \to a_{i-j} \ea
\label{fusionring}\ee
where the choice between the first two is determined by which index lies
in the range of physical particles $1\dots,p$.  These fusion
properties will play a key role in the resolution of the form--factor
equations.

The paper is organized as follows: In the next section we give a short
review of the form--factor axioms for models with only scalar
particles. In section \ref{sec-para} we parametrize the form--factors
in a way to reduce the resolution of the axioms to polynomial
recursion relations. The Watson's equations are fulfilled
automatically by this parametrization and the kinematical pole
equation (\ref{kin}) reduces to that encountered in the
Sinh--Gordon model \cite{FMS,a10}. In section \ref{sec-examples} some
examples are discussed, namely the theories $M_{2,5}$ (the Yang--Lee
model) and the two-particle system $\M_{2,7}$. We determine the
operator--content of these models by examining the bound-state axiom
(\ref{bounds}). These methods are generalized in section
\ref{sec-all} to the whole series $\M_{2,2p+3}$. The key result of the
paper is that the bound state axiom reduces for these models to a {\em
single} polynomial recursion relation for the form--factors involving
only particles of type one.  This recursion relation is explicitly
calculated and the dimensionality of the solution-space determined.
This also gives the operator content of these models. We give the
physical interpretation of these operators. An account on descendent
operators is given in section \ref{sec-desc} and finally in section
\ref{sec-conc} we present our conclusions.

\resection{Form--Factor Axioms for Systems with Scalar Particles
\label{sec-ff}}

Let us review the properties that the form--factors must satisfy.
We discuss them for the case that the spectrum consists of only scalar
self-conjugated particles.
They derive from crossing symmetry, CPT invariance and the properties
of the Faddeev- Zamolodchikov operators
\cite{Smirnov1,Karowski,nankai}.
 The physical vacuum $\vert 0 \rangle$ is annihilated by operators
$Z^{\epsilon} (\beta)$, $$ Z^{\epsilon}(\beta) \vert 0 \rangle=0 \sb ,
$$ and the physical states are created by
\be \vert Z_{\epsilon_1}(\beta_1)\dots Z_{\epsilon_n}(\beta_n) \rangle =
Z_{\epsilon_1}^\star (\beta_1) \dots Z_{\epsilon_n}^\star (\beta_n)
\vert 0 \rangle \sb .
\label{zstates} \ee
The importance of these operators lies in their commutation relations,
which are governed by the $S$-matrix,
\bea
Z^{\epsilon_1}(\beta_1) Z^{\epsilon_2}(\beta_2) &=&
S_{\epsilon_1\epsilon_2}(\beta_1-\beta_2)
Z^{\epsilon_2}(\beta_2) Z^{\epsilon_1}(\beta_1)
\sb ,\nonumber\\
Z_{\epsilon_1}^\star(\beta_1) Z_{\epsilon_2}^\star(\beta_2) &=&
S_{\epsilon_1\epsilon_2}(\beta_1-\beta_2)
Z_{\epsilon_2}^\star(\beta_2) Z_{\epsilon_1}^\star(\beta_1)
\sb ,\label{s-comm}\\
Z^{\epsilon_1}(\beta_1) Z_{\epsilon_2}^\star(\beta_2) &=&
Z_{\epsilon_2}^\star(\beta_2)
S_{\epsilon_1\epsilon_2}(\beta_1-\beta_2)
Z^{\epsilon_1}(\beta_1) +2 \pi \delta_{\epsilon_2}^{\epsilon_1}
\delta(\beta_1-\beta_2)
\sb .\nonumber\eea

A consequence of the commutation relations (\ref{s-comm}) is the
following symmetry property of the form--factors,
\bea
& & F^\O_{\e_1\dots\e_i\e_{i+1}\dots\e_n}
(\beta_1,\dots,\beta_i,\beta_{i+1},\dots\beta_n) =\nonumber\\ &
&S_{\e_i\e_{i+1}}(\beta_i-\beta_{i+1})\,
F^\O_{\e_1\dots\e_{i+1}\e_{i}\dots\e_n}
(\beta_1,\dots,\beta_{i+1},\beta_{i},\dots\beta_n) \sb .
\label{wat1}\eea
Consider the analytic continuation $\beta_1 \goto \beta_1+2\pi i$,
which from the kinematical point of view brings back to the initial
configuration, but changes the ordering of the particles in the
function $ F^\O_{\e_1\dots\e_n} (\beta_1,\dots,\beta_n )$.  This
analytic continuation can be related to the original form--factor in
an alternative way by scattering all other particles.  The consequence
for the form--factor is the constraint equation
\bea & &
F^\O_{\e_1\e_2\dots\e_n} (\beta_1+2\pi i,\beta_2,\dots,\beta_n )=
F^\O_{\e_2\dots\e_n\e_1} (\beta_2,\dots,\beta_n,\beta_1 )=\nonumber\\
& &S_{\e_1\e_2} S_{\e_1\e_3}
\dots
S_{\e_1\e_n}\,\, F^\O_{\e_1\e_2\dots\e_n}
(\beta_1,\beta_2,\dots,\beta_n )
\sb .\label{wat2}\eea

A further constraint is imposed by relativistic invariance. Assume
that the operator $\O$ has spin $s$. Then
\be
 F^\O_{\e_1\dots\e_n} (\beta_1+\Lambda,\dots,\beta_n+\Lambda ) =
e^{s\Lambda} F^\O_{\e_1\dots\e_n} (\beta_1,\dots,\beta_n )
\sb .\label{prop2}\ee

The form-factor equations discussed so far connect form-factors
corresponding to a fixed particle number.  $n$ and $m$ particle
form--factors are still independent of each other.  The following two
constraint equations have a recursive structure and link form--factors
of {\em different} particle numbers. They originate from the
pole--structure of the form--factors.

The first type of poles has a kinematical origin and corresponds to
zero--angle scattering, $$ -i \lim_{\beta'\to \beta}(\beta'-\beta)
F^\O_{\e\e\e_1\dots\e_n} (\beta'+i\pi,\beta,\beta_1,\dots,\beta_n) =$$
\nopagebreak
\be\left (1-\prod_{i=1}^{n} S_{\e\e_i}(\beta-\beta_i) \right )
F^\O_{\e_1\dots\e_n} (\beta_1,\dots,\beta_n )
\sb .\label{kin} \ee

If particles $A_i$, $A_j$ form a bound state $A_k$, the corresponding
two-particle scattering amplitude exhibits a pole with the residue
\be
 -i \lim_{\beta'\to iu_{\e_i\e_j}^{\e_k}}(\beta-iu_{\e_i\e_j}^{\e_k})
S_{\e_i\e_j}(\beta) =(\Gamma_{\e_i\e_j}^{\e_k})^2 \sb ;
\label{residueS}\ee
$\Gamma_{\e_i\e_j}^{\e_k}$ is the three--particle on--shell vertex.
Corresponding to this bound state the form--factor exhibits a pole at
the point $ \beta_i-\beta_j = i \bar u_{\e_i\e_j}^{\e_k} $ with the
residue $$ -i \lim_{\beta'\to \beta}(\beta'-\beta)
F^\O_{\e_1\dots\e_i\e_j\dots\e_n} (\beta_1,\dots,\beta'+i\bar
u_{\e_i\e_k}^{\e_j},\beta-i\bar u_{\e_j\e_k}^
{\e_i},\dots,\beta_{n-1}) = $$
\be =\Gamma_{\e_i\e_j}^{\e_k} F^\O_{\e_1\dots\e_k\dots\e_n}
 (\beta_1,\dots,\beta,\dots,\beta_{n-1} )
\sb . \label{bounds}\ee

The equations (\ref{wat1})--(\ref{prop2}) together with the residue
relations (\ref{bounds}) and (\ref{kin}) can be used as a system of
axioms for the form-factors. In order for the operators to satisfy
proper locality relations their form-factors have to satisfy the following
ultraviolet bound
\cite{nankai}
\be
F_{\e_1,\dots,\e_n}(\beta_1+\Lambda,\dots,\beta_i +\Lambda,
\beta_{i+1}, \dots , \beta_{n}) = O(e^{S |\Lambda |}) \sb {\rm for}
\sb |\Lambda| \sim \infty \sb ,
\label{asym}\ee
where $S$ is a number common for all $i$ and $n$.

\resection{Parametrization of the $n$--Particle Form--Factor \label{sec-para}}

In order to find solutions of the above discussed equations one needs
to find a convenient parametrization of the form--factors.  A solution
process which has proved to be very useful \cite{FMS,Karowski,YLZam} is
to start with the calculation of the two--particle form--factor and
then to parametrize the $n$--particle form--factor in terms of it.
Let us discuss these steps in detail.

The Watson's equations for $n=2$ read as
\be F^\O_{ab} (\beta) = S_{ab} (\beta) F^\O_{ba} (-\beta) \sa,\sb
F^\O_{ab}(i\pi - \beta) = F^\O_{ba}(i \pi+\beta) \sb .\label{w2}\ee
This set of equations can be solved with the help of the following
observation
\cite{Karowski}. If the $S$--matrix element $S_{ab}$ can be written in an
integral representation of the form
\be S_{ab}(\beta) = \exp \left \{ \int_{0}^{\infty} \frac{dx}x f(x)
\sinh \left ( \frac{x\beta}{i\pi}\right ) \right \} \sb ,\ee
then a solution of (\ref{w2}) is given by
\be F^\O_{ab}(\beta) = \exp \left \{ \int_{0}^{\infty} \frac{dx}x f(x)
\frac{
\sin^2 \left ( \frac{x(i\pi-\beta)}{2\pi}\right )}{
\sinh x} \right \} \sb .\label{exprr}\ee
Note that multiplying the expression (\ref{exprr}) by an arbitrary
function of $\cosh \beta$ we find another solution of equations
(\ref{w2}).  In order to determine the final form of $F_{ab}(\beta)$
it is necessary to consider a specific theory and to know the physical
nature of the operator $\O$.  In the following we drop the index
referring to the operator $\O$ keeping this ambiguity in mind.

 In order to select one specific solution we define the {\em minimal
two particle form--factor} $F_{ab}^{\rm min}$ as the solution of equations
(\ref{w2}) with the additional property that it is analytic in $0<{\rm
Im} \beta < \pi$ and has no zeros in this range.

We choose for simplicity the form--factor $F_{1,\dots,1}$ where the
indices indicate that we discuss the form--factor corresponding to the
fundamental particle of the considered theory. This is just a
technical simplification since other form--factors can be treated in a
similar way.  Moreover we will see that all other form--factors can be
obtained from this one by means of the bound state equation
(\ref{bounds}).  For this we need only $F_{11}^{\rm min}$
\cite{Karowski}
\be
F^{\rm min}_{11}(\beta) = (-i) \sinh \frac\beta 2 \, \exp \left \{
\int_0^\infty \frac{dx}x \frac{\cosh x (\frac 12 - \frac\alpha\pi)}
{\cosh \frac x2}
\frac{\sin ^2 \frac{ x(i \pi-\beta)}{2\pi}}{\sinh x} \right \}
\sb .\ee
It is useful to rewrite this expression in terms of
$\Gamma$--functions as
\be
F^{\rm min}_{11}(\beta) = (-i) \sinh \frac\beta 2 \,\frac{\zeta(\beta)}{\zeta(i
\pi)}
\sb ,\ee
with $$
\zeta (\beta) =\prod_{k=0}^\infty
\frac{ \Gamma\left (k+\frac 12 +\frac \alpha {2\pi} -\frac{i \beta}2 \right )
\Gamma\left (k+1 -\frac \alpha {2\pi} -\frac{i \beta}2 \right )}{
\Gamma\left (k+\frac 12 -\frac \alpha{ 2\pi} -\frac{i \beta}2 \right )
\Gamma\left (k+\frac \alpha{ 2\pi} -\frac{i \beta}2 \right )}
\frac
{\Gamma\left (k+\frac 32+ \frac \alpha {2\pi} +\frac{i \beta}2 \right
)
\Gamma\left (k+2- \frac \alpha {2\pi} +\frac{i \beta}2 \right )
}{\Gamma\left (k+\frac 32 -\frac \alpha {2\pi} +\frac{i \beta}2 \right
)
\Gamma\left (k+1+ \frac \alpha {2\pi} +\frac{i \beta}2 \right )
}$$ For later convienience we also give the explicit expression for
the constant $\zeta(i\pi)$
\bea \zeta (i \pi) &=& \prod_{k=0}^\infty \left ( \frac{
\Gamma(k+1+\frac \alpha {2\pi} )
\Gamma(k+\frac 32-\frac \alpha {2\pi} )
}{\Gamma(k+1-\frac \alpha {2\pi} )
\Gamma(k+\frac 12+\frac \alpha {2\pi} )
}\right )^2 = \\ &=&
\frac 1 \pi \exp \left \{ 4 \int dt \frac{\sinh \frac t2 \sinh \frac
\alpha {2\pi} t \sinh \frac 12 (1-\frac\alpha\pi)  t }{t \sinh^2 t} \right \}
\eea
 In the following we will drop the indices referring to particle $1$
and denote it simply as
$$F_{\underbrace{1,1,\dots,1}_n}(\beta_1,\dots,\beta_n) \equiv
F_n(\beta_1,\dots,\beta_n) \sb .$$

In general this form--factor can be parametrized as \cite{Karowski}
\be F_n (\beta_1,\dots,\beta_n) =K_n(\beta_1,\dots,\beta_n) \prod_{i<j}
F^{\rm min}(\beta_{ij}) \sb ,
\ee
where the function $K_n$ needs to satisfy Watson's equations
(\ref{wat1}) and (\ref{wat2}) with an $S$--matrix factor $S=1$.
Therefore it is a completely symmetric, $2 \pi i$--periodic function
of $\beta_i$. It must contain all expected kinematical and bound state
poles.  Finally it will contain the information on the operator $\O$.

Since we know the possible scattering processes we can split the
function $K_n$ further in order to determine the pole structure. The
kinematical poles are expected at the rapidity values $\beta_i \to
\beta_j +i \pi$.  These poles can be generated by the completely
symmetric function $\prod_{i<j}(x_i+x_j)^{-1}$, where we have
introduced the notation $x_i \equiv e^{\beta_i}$.  Further, the
$S$--matrix element $S_{11}$ exhibits a pole at $\beta = i \alpha $.
Then according to (\ref{bounds}) also the form--factor exhibits poles
which can be generated by the function $$ \frac 1 { \sinh \frac 12 (
\beta_{ij} - i \alpha ) \sinh \frac 12 ( \beta_{ij} + i \alpha )} \sb
.$$

The final parametrization of the $n$--particle form--factor reads as
\be F_n (\beta_1,\dots,\beta_n) = \tilde{Q}_n(\beta_1,\dots,\beta_n)
 \prod_{i<j} \frac{F^{\rm min}(\beta_{ij})}{ (x_i+x_j) \sinh \frac 12 (
\beta_{ij} - i \alpha ) \sinh \frac 12 ( \beta_{ij} + i \alpha )} \sb
,\label{paraf}\ee $\tilde{Q}_n$ is now a symmetric function free of
singularities.  The form factor equations have been reduced through
this parametrization to a set of coupled recursive relations for
$\tilde{Q}_n$.

Using the parametrization (\ref{paraf}) it is straightforward to
derive the recursion relation corresponding to the kinematical poles.
Let us introduce $\omega=e^{i \alpha}$. We find
\be \tilde{Q}_{n+2} (-x,x,x_1,\dots,x_n) = D_n(x,x_1,\dots,x_n)
\tilde{Q}_n (x_1,\dots ,x_n) \label{recursive}\ee
with $$D_n(x,x_1,\dots,x_n) = (-1)^{n+1} (\zeta(i \pi) \pi)^{2n} \cos
^2 (\frac \alpha 2) $$ $$\times
\left (\prod_{i=1}^{n}(x \omega^{\frac 12} +x_i  \omega^{-\frac 12})
(x \omega^{-\frac 12} -x_i \omega^{\frac 12}) -\prod_{i=1}^n(x
\omega^{\frac 12} -x_i \omega^{-\frac 12}) (x \omega^{-\frac 12} +x_i
\omega^{\frac 12})\right )
\sb .$$
It is useful to define
\be\tilde{Q}_n = H_n Q_n \label{para}\sb ,\ee
with
\be
H_n = C_0 \left ( 4 \cos ^2 \frac \alpha 2\, \sin \alpha \right )^{
\frac n 2} (\zeta(i\pi)\pi)^{\frac{(n-1)^2-1}{2}}\, i^{n^2}
\sb .\label{hn}\ee
In this way the recursion relation for $Q_n$ coincides exactly with
that of the Sinh--Gordon model \cite{FMS,a10}.

Before going on to the task of actually solving the form-factor
equations let us discuss the structure of the functions
$Q_n(x_1,\dots,x_n)$. We require them to be fully symmetric homogenous
functions, analytic apart from the origin and polynomially bounded
because of (\ref{asym}). Further from Lorentz invariance it follows
that $Q_1 (x_1) = const.$.  Therefore $Q_n$ has to take the structure
\be Q_n(x_1,\dots,x_n) = \sum_N \frac{1}{\prod_{i=1}^nx_i^N} \P (x_1,
\dots,x_n) \sb , \label{ansatz}
\ee
where the sum needs to be finite and $\P$ are polynomials whose degree
is fixed by Lorentz invariance.  We choose as a basis in this space
the single terms in this sum, determined by a fixed integer $N$.

Let us discuss this ansatz. It results from a strict interpretation of
the form-factor axioms. The problem is that in this way it seems
difficult to obtain a direct correspondence with operators in the
UV-limit conformal field theory. This is not the case for primary
fields as we will see, but for descendent operators one cannot expect
such a simple structure. This is due to the mixing of operators of
{\em different } Verma modules in the perturbed theory. This occurs
for example between operators of the Verma-module of the Identity and
of the Verma module of the perturbing field, which constitute the
conservation laws in the perturbed model \cite{Zam}.

This problem was analyzed in  \cite{Christe} for the example of the
Yang-Lee model. In order to get a direct correspondance
with the descendents of the field $\phi_{13}$ the admixture of the
operators from the module of the identity had to be taken in account.
This led to the introduction of further singularities.

In this article we investigate mainly the form--factors of what we will
call primary operators. This notation is borrowed from the
corresponding conformal field theories, where the operators are
organized in a Verma module structure.  The scalar operator with the
lowest scaling dimension is identified as the primary field.
Analogously we define as {\em primary operators in the perturbed
model} those spinless fields which have the mildest ultraviolet
behaviour. The scaling dimension increases with $N$ and therefore we
define as primary operators those with $N=0$.

In section \ref{sec-desc} we will discuss the problem of the
descendent operators using the ansatz (\ref{ansatz}). Inspite of the
problems discussed we will see a remarkable coincidence with the
structure of the conformal field theories.

\subsection{Resolution of the kinematical Recursion Relation
 \label{sec-SHG}}

Since $Q_n(x_1,\dots,x_n)$ is a symmetric polynomial satisfying
(\ref{recursive}) it is useful to introduce as a basis in this space
the {\em elementary symmetric polynomials} $\sigma^{(n)}_k(x_1, \dots
, x_n)$ which are generated by
\cite{Macdon}
\EQ
\prod_{i=1}^n(x+x_i)\,=\,
\sum_{k=0}^n x^{n-k} \,\sigma_k^{(n)}(x_1,x_2,\ldots,x_n).
\label{generating}
\EN
Conventionally the $\sigma_k^{(n)}$ with $k>n$ and with $n<0$ are
zero.  The explicit expressions for the other cases are
\EQ
\begin{array}{l}
\sigma_0=1\hs\hs,\\
\sigma_1=x_1+x_2+\ldots +x_n\hs\hs,\\
\sigma_2=x_1x_2+x_1x_3+\ldots x_{n-1}x_n\hs\hs,\\
\vdots  \qquad \qquad \vdots  \\
\sigma_n=x_1x_2\ldots x_n\hs\hs\hs.
\end{array}
\EN
The $\sigma_k^{(n)}$ are linear in each variable $x_i$ and their total
degree is $k$.

In terms of this basis the recursive equations (\ref{recursive}) take
the form
\EQ
(-1)^n\,Q_{n+2}(-x,x,x_1,\ldots,x_n)\, = \,x D_n(x,x_1,x_2,\ldots
,x_n)
\,Q_n(x_1,x_2,\ldots,x_n)
\label{rec}
\EN
where
\be D_n(x,x_1,\dots,x_n) = \sum_{k=1}^n \sum_{m=1,odd}^k [m]\, x^{2(n-k)+m}
\sigma_{k}^{(n)}\sigma_{k-m}^{(n)} (-1)^{k+1} \sb .
\label{D_n}
\ee
We have introduced the symbol $[l]$ defined by
\be
[l]\equiv\frac{\sin ( l\alpha)}{\sin\alpha}
\ee

The recursion relation (\ref{rec}) was solved in the space of
polynomials in
\cite{a10}.  There a class of solutions was found, given by
\be
Q_{n}(k) = \vert\vert M_{ij}(k) \vert\vert \sb ,
\ee
where $M_{ij}(k)$ is an $(n-1)\times (n-1)$ matrix with entries
\be M_{ij}(k) =
\s_{2i-j}\, [i-j+k] \sb .
\label{element}
\ee
In Sinh-Gordon theory the operators corresponding to these solutions
are the exponentials $e^{kg\phi}$, g being the coupling constant and
$\phi$ the elementary field appearing in the Lagrangian.

Important properties of the polynomials $Q_n$ can be obtained by
analyzing the recursive equations (\ref{rec}) \cite{a10}.  We are
interested in the dimensionality of the solution space of this
recursion relation. It is given by $\dim (Q_{2n-1}) = \dim (Q_{2n})
=n$. The proof is done by induction: Lorentz invariance (\ref{prop2})
fixes the total degree of the polynomials $Q_n$ as $\frac 12 n (n-1)$.
This implies that $Q_1 = A_1$ and $Q_2 = A_2\s_1$, with $A_1,A_2$
arbitrary constants. Let us examine the polynomials $Q_n$ with odd
index $n$. Assume that $\dim (Q_{2n-1}) =n$.  Then the
dimensionality of $Q_{2n+1}$ is given by $\dim (Q_{2n-1})$ plus the
dimension of the kernel of the recursion relation {\em i.e.} by
\EQ
Q_{2n+1}(-x,x,\dots,x_{n+2}) =0 \,\,\, .
\EN
In the space of polynomials ${\P}$ of total degree $(2n+1)n$,
there is only one solution of this equation,
\EQ
Q_{2n+1} = \prod_{i < j}^{2n+1} (x_i + x_j) \,\,\, .
\label{kernel}
\EN
This polynomial has partial degree $2n$ and coincides with the
denominator in eq.\,(\ref{para}).

Since the kernel is a one-dimensional manifold, the dimension of the
space of solutions increases exactly by one at each step of the
recursion, and we have proved our statement above. The proof goes
analogously for the polynomials with even index.

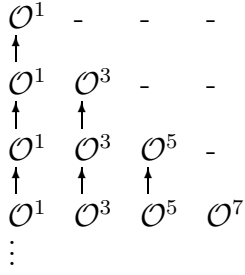
\begin{figure}
$$
\begin{picture}(100,115)(0,-100)
\put(0,0){$\O^1$}
\put(25,0){-}
\put(50,0){-}
\put(75,0){-}
\put(3,-13){\vector(0,1){10}}
\put(0,-25){$\O^1$}
\put(25,-25){$\O^3$}
\put(50,-25){-}
\put(75,-25){-}
\put(0,-50){$\O^1$}
\put(3,-38){\vector(0,1){10}}
\put(25,-50){$\O^3$}
\put(28,-38){\vector(0,1){10}}
\put(50,-50){$\O^5$}
\put(75,-50){-}
\put(0,-75){$\O^1$}
\put(3,-63){\vector(0,1){10}}
\put(25,-75){$\O^3$}
\put(28,-63){\vector(0,1){10}}
\put(50,-75){$\O^5$}
\put(53,-63){\vector(0,1){10}}
\put(75,-75){$\O^7$}
\put(0,-90){$\vdots$}
\end{picture}
$$
\caption{Tower like structure of the operators $\O^n$, determined by
the independent solutions of the kinematical recursion relation}
\label{fig-f1}
\end{figure}

In figure \ref{fig-f1} we have exhibited the structure of the solution
space. It has a tower-like structure growing linearly with every step
of the kinematical recursion relation. It will be interesting to see
how this tower gets truncated by the additional implementation of the
bound state axiom (\ref{bounds}).

Note that the polynomials $Q_n(k)$, with $k$ integer do not form a
base in this space.  This is because of their periodic dependence on
the coupling $\alpha$, so that for $\M_{2,2p+3}$ only the
first $p$ polynomials are independent. We will see in the following
how by enforcing the bound state recursion relation
(\ref{bounds}) the solution space gets restricted such that the
$Q_n(k)$, $k=1,\dots,p$ will form a base.

\resection{Examples\label{sec-examples}}

\subsection{The Yang--Lee Model}

The form--factors of the Yang--Lee model have been analyzed
extensively
\cite{Christe,smirnov13,YLZam}. We want to use it here as an example in
order to introduce the techniques which we will generalize in the
following sections.

For the Yang--Lee model, $M_{2,5}$ $\alpha= \frac 23 \pi$ and the
theory contains only one massive particle. It is for this reason not a
typical example for the series $M_{2, 2p+3}$. Nevertheless let us see
how the solution space of the recursive equations can be classified.
\
The recursion relation for the bound state equation (\ref{bounds})
reads as
\cite{YLZam}
\be
Q_{n+1}(x\omega^{\frac 12}, x\omega^{-\frac 12},x_1,\dots,x_{n-1}) = x
\prod_{i=1}^{n-1} (x+x_i) Q_n(x,x_1,\dots,x_{n-1})
\sb .\label{bsyl}\ee

We saw that the solution space for the kinematical recursion relation
was linearly growing. This is not the case if we impose {\em both}
recursion relations, since the kernel of the kinematical recursion
relation is not a solution of the bound state equation in a polynomial
space.

Let us analyze the lowest polynomials $Q_n$. From the solution of the
kinematical recursion relation we know that the first polynomials
$Q_1$ to $Q_3$ have the general form $$ Q_1 =A_1 \sb ,\sa Q_2
=A_2\s_1 \sb ,\sa Q_3= A_3
\s_3 +(A_1-A_3) \s_2 \s_1 \sb ,$$
with arbitrary constants $A_1,A_2,A_3$. Now we additionally impose the
bound state recursive equation (\ref{bsyl}) from which follows after a
short calculation that
the only consistent solution is $$ Q_1 =A_1 \sb , \sa Q_2 =A_1\s_1 \sb
, \sa Q_3= A_1 \s_2 \s_1 \sb .$$ In terms of the elementary solutions
introduced in the last section this solution corresponds to $Q_n(1)$.

In general we have to show that the kernel solutions of the
kinematical recursive equation (\ref{kernel}) are not solutions to the
bound state equation. Suppose $Q_i =0$, for $i=1,\dots,n$.  Then any
non--zero solution for $Q_{n+1}$ must have zeros as well at locations
of the kinematical poles as at those of the bound state poles in order
to give a zero residue. This is, that
\be Q_{n+1} \sim \K_{n+1}=
 \prod_{i<j}^{n+1} (x_i+x_j)(x_i+\omega x_j) (x_i+\omega
^{-1} x_j) \sb .\label{above}\ee On the other hand we know that
because of relativistic invariance the total degree of $Q_{n+1}$ must
be $\frac{(n+1)n}2$. The above expression (\ref{above}) has total
degree $\frac{3(n+1)n}2$, which shows that it cannot be a solution of
the combined recursion relations.  That is, the Kernel of the combined
recursion relations (\ref{rec}) and (\ref{bsyl}) is zero-dimensional.
Therefore {\em no additional} solutions exist to $Q_n(1)$, which was
already found in \cite{YLZam}. The operator corresponding to this
specific solution is the trace of the stress energy tensor.

This simple argument has a rather important physical consequence. It
shows that also for the perturbed Yang--Lee model the space of primary
operators (which in our case corresponds to polynomial solutions of
the recursive equations) is one--dimensional, as it is the case for
the conformal field theory.

\subsection{The Model $M_{2,7}$}

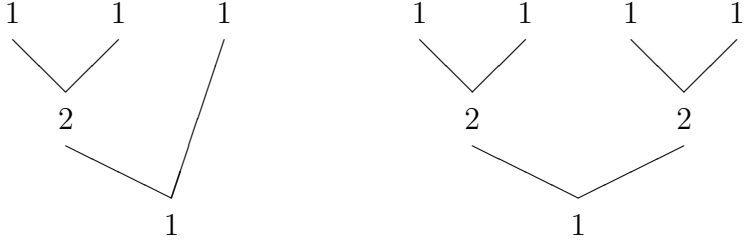
\begin{figure}
\begin{picture}(150,100)
\put(40,100){\makebox(0,0){1}}
\put(80,100){\makebox(0,0){1}}
\put(120,100){\makebox(0,0){1}}
\put(40,90){\line(1,-1){20}}
\put(80,90){\line(-1,-1){20}}
\put(60,60){\makebox(0,0){2}}
\put(60,50){\line(2,-1){40}}
\put(120,90){\line(-1,-3){20}}
\put(100,20){\makebox(0,0){1}}
\end{picture}
\begin{picture}(150,100)
\put(40,100){\makebox(0,0){1}}
\put(80,100){\makebox(0,0){1}}
\put(120,100){\makebox(0,0){1}}
\put(160,100){\makebox(0,0){1}}
\put(40,90){\line(1,-1){20}}
\put(80,90){\line(-1,-1){20}}
\put(120,90){\line(1,-1){20}}
\put(160,90){\line(-1,-1){20}}
\put(60,60){\makebox(0,0){2}}
\put(140,60){\makebox(0,0){2}}
\put(60,50){\line(2,-1){40}}
\put(140,50){\line(-2,-1){40}}
\put(100,20){\makebox(0,0){1}}
\end{picture}
\caption{Possible fusion processes leading back to particle 1}
\label{fig-exam}
\end{figure}

As a next step we examine the two--particle system, defined by the
perturbed $M_{2,7}$ theory.  The $S$--matrix of this model is given by
\be S_{11}=f_{\frac 25\pi} \sb , \sa S_{12}=f_{\frac 35\pi}f_{\frac 15\pi}
\sb ,\sa S_{22}=(f_{\frac 25\pi})^2 f_{\frac 15\pi}\sb .
\ee
The corresponding fusion angles are
\be u_{11}^2 = \frac{2 \pi} 5\sb ,\sa
 u_{12}^1 = \frac{4 \pi} 5
\sb ,\sa u_{12}^2 = \frac{3 \pi} 5
\sb ,\sa u_{22}^1 = \frac{4 \pi} 5 \sb .\ee
Now we need to analyze all form--factors $F_{\e_1,\e_2\dots\e_n}$
where $\e_i$ can take the values 1 or 2.  By the bound state equation
(\ref{bounds}) any form--factor containing the particle 2 can be
expressed in terms of residues of form--factors containing only
particle 1.  This reflects the situation in $S$--matrix theory where
it suffices to know the $S$--matrix of the fundamental particle in
order to calculate the full $S$--matrix. Therefore we concentrate on
the form--factor with all indices corresponding to the particle 1. The
difficulty is now that the bound--state residue equation not only
links the different form--factors but gives also constraints on the
form--factors of the particle 1. This since we can return by various
fusions to the particle 1. Two examples are shown in figure
\ref{fig-exam}.

Let us analyze the first of these processes. If one uses two times the
bound state equation the corresponding recursion relation is $$ (-1)
\lim_{\tilde{\beta} \to \beta} \lim_{\hat{\beta} \to \beta}
(\tilde{\beta} -\beta)(\hat{\beta}-\beta) F_{n+2} (\hat{\beta}+2i \bar
{u}_{12}^1,\tilde \beta,\beta-i \bar{u}_{11}^2,
\beta_1,\dots,\beta_{n-1}) =
$$
\be
=\Gamma_{11}^2 \Gamma_{21}^1 F_n(\beta,\beta_1,\dots,\beta_{n-1}) \sb .\ee
Consider the difference between the first and third rapidity in this
formula $$\tilde{\beta}+2i\bar{u}_{12}^1 -\beta+i\bar{u}_{11}^2 =
\tilde{\beta}-\beta +i\pi  \sb .$$
The corresponding recursive equation for the $Q_n$ will be a special
case of the kinematical recursion relation, and therefore this process
does not give any new constraint on the solutions.

The second process in figure \ref{fig-exam} on the other hand gives an
independent constraint. It consists of three fusion processes and the
corresponding form--factor equation reads as $$(-i)^3
\lim_{\hat{\beta}\to\tilde{\beta}}
\lim_{\beta'\to\beta}\lim_{\tilde{\beta}\to\beta}
(\hat{\beta}-\tilde{\beta})(\beta'-\beta) (\tilde{\beta}-\beta)
F_{n+3}(\hat{\beta}+\frac{3i\pi}5,\tilde{\beta}+\frac{i\pi}5,
\beta '-\frac{i\pi}5,\beta-\frac{3i\pi}5,\beta_1,\dots,\beta_{n-1})  $$
\be = (\Gamma_{11}^2)^2 \,\Gamma_{22}^1\, F_n(\beta,\beta_1,\dots,\beta_{n-1})
\sb .\label{b27}\ee
Note that in this relation no rapidity difference of $i\pi$ appears
and therefore this relation is independent from the kinematical
recursion relation. All other possible fusion processes
can be reduced either to (\ref{b27}) or
to the kinematical residue equation (\ref{rec}).

For this model, containing only two particles this statement is easy
to understand: In order to be sure that we have obtained all possible
constraint equations, we need to consider all possible fusion
processes which return to particle 1. Considering a specific fusion
process it is not important which set of particles will be fused. The
resulting relation for $Q_n$ will always be the same, because $Q_n$
is a symmetric polynomial.

Further one needs to consider only elementary fusion blocks, {\em
i.e.} processes which come back to particle 1 only once, since other
processes can be obtained by combining elementary ones.

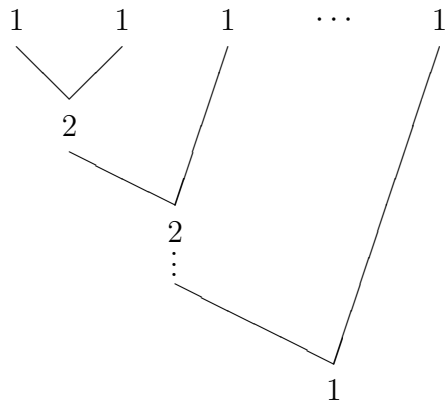
\begin{figure}[th]
\begin{picture}(150,130)(-40,-30)
\put(40,100){\makebox(0,0){1}}
\put(80,100){\makebox(0,0){1}}
\put(120,100){\makebox(0,0){1}}
\put(160,100){\makebox(0,0){$\cdots$}}
\put(200,100){\makebox(0,0){1}}
\put(40,90){\line(1,-1){20}}
\put(80,90){\line(-1,-1){20}}
\put(60,60){\makebox(0,0){2}}
\put(60,50){\line(2,-1){40}}
\put(120,90){\line(-1,-3){20}}
\put(100,20){\makebox(0,0){2}}
\put(98,3){$\vdots$}
\put(200,90){\line(-1,-3){40}}
\put(160,-30){\line(-2,1){60}}
\put(160,-40){\makebox(0,0){$1$}}
\end{picture}
\caption{Process of multiple application of the fusion
$1 \times 2 \to 2$}
\label{fig-exam2}
\end{figure}

Now in the model $\M_{2,7}$ we have only two particles and there are
only few elementary fusion processes. In addition to the ones shown in
fig. \ref{fig-exam} the only other type of processes are those
resulting from an multiple additional application of the fusion
$1\times 2 \to 2$. One example is shown in fig. \ref{fig-exam2}.
Analyzing the rapidity shifts entering the recursion relation, one
will see immediately that any such process can be decomposed in
applications of the recursion relations (\ref{kin}) and (\ref{b27}).

The next step is therefore to calculate the corresponding recursion
relation for $Q_n$. It is given by
\be Q_{n+3} (x\omega^{\frac 32},x \omega^{\frac 12},x \omega^{-\frac 12}
,x \omega^{-\frac 32},x_1,\dots,x_{n-1} ) =
U_n(x,x_1,\dots,x_{n-1})Q_n(x,x_1,\dots,x_{n-1})
\sb ,\label{bs27}\ee
with
\be
U_n(x,x_1,\dots,x_{n-1}) = x^6 \prod_{i=1}^{n-1}
(x+x_i)(x-x_i\omega^2)(x-x_i\omega^{-2})
\sb .\label{u27}\ee

As in the Yang--Lee model we have two polynomial equations to solve.
The relation (\ref{bs27}) again couples the odd and even form--factors
but now the form--factor $F_{n+3}$ and $F_n$, whereas in the Yang--Lee
model the recursion relation (\ref{bsyl}) related $F_{n+1}$ and $F_n$.

Let us discuss now the dimension of the solution--space of the coupled
recursive equations. First note that also in this case the kernel
solutions of the kinematical recursive equation in general cannot be
solutions to the bound state equation. The argument is similar as for
the Yang--Lee model. We present it here in a slightly different
form.

Suppose the solution--space has a certain dimensionality at level $n$.
Examine the kernel of the kinematical recursion relation $$\K_{n+2}^{kin} =
\prod_{i<j}^{n+2} (x_i+x_j)\sb .$$
If this is a solution of the bound
state equation then $\K_{n+2}^{kin} \to U_{n- 1} Q_{n-1}$. That is,
$Q_{n+2}^K$ must factor into $U_{n-1}$ times a polynomial. Now,
$$\K_{n+2}^{kin} (x\omega^{\frac 32},x \omega^{\frac 12},x \omega^{-\frac
12} ,x \omega^{-\frac 32},x_1,\dots,x_{n-1} ) = $$
$$x^6\prod_{i=1}^{n-2} (x \omega^{\frac 32} +x_i) (x \omega^{\frac 12}
+x_i)(x \omega^{-\frac 12} +x_i) (x \omega^{-\frac 32} +x_i)
\prod_{i<j}^{n-2}(x_i+x_j) $$
Factoring out $U_{n-1}$ $\K_{n+2}^{kin}$ takes the form $$\K_{n+2}^{kin}
(x\omega^{\frac 32},x \omega^{\frac 12},x \omega^{-\frac 12} ,x
\omega^{-\frac 32},x_1,\dots,x_{n-1} ) = $$
$$U_{n-1}(x,x_1,\dots,x_{n-2}) \frac{(-x \omega+x_i) (-x\omega
+x_i)}{(x+x_i)}
\prod_{i<j}^{n-2}(x_i+x_j) $$
Since the terms containing $x$ cannot cancel for $\omega=e^{\frac{2 i
\pi}5}$ the right hand side is not of polynomial form, and therefore
$\K_{n+2}^{kin}$ is not a solution of the bound state recursion
relation, in the space we are considering.

This argument obviously works only for $n>4$. This means that the
dimensionality of the combined solution space is fixed by that of the
first 4 polynomials $Q_1,
\dots Q_4 $. Let us analyze them. The first three of them $Q_1,Q_2,Q_3$
remain untouched by the bound state recursion relation and therefore
are determined by the kinematical recursion relation and have
dimensionality $1,1,2$ respectively. The last one links the even and
odd sectors of the polynomial, but for any solution $Q_4$ of the
kinematical recursion relation we can choose a constant $Q_1$ such
that the bound state recursion relation is fulfilled. Therefore the
solution space of $Q_4$ is also two--dimensional.

With this analysis we have shown that the solution space for the
recursion relations of the perturbed model $\M_{2,7}$ is
two--dimensional. This implies that also the perturbed model contains
only two primary operators as the conformal one.  A base in this space
is given by the elementary solutions $Q(1)$ and $Q(2)$. A physical
interpretation will be given in section
\ref{sec-six}.

\resection{The Models $\M_{2,2p+3}$\label{sec-all}}

In the last section we discussed how in the two--particle system
$\M_{2,7}$ the form--factor constraints have been reduced to only {\em
two} polynomial recursion equations, the kinematical one (\ref{rec})
and one equation deriving from a multiple application of the bound
state form factor axiom (\ref{b27}). We will now show, that this is
true for all models of the series $\M_{2,2p+3}$, which contain $p$
particles.

Let us now describe the key point of how it is possible to resolve the
form--factor axioms for a system containing more than one particle.
The difficulty lies in the fact that in general form--factors
corresponding to different particles will be linked by the bound state
equations. Therefore one has not to resolve only one equation but a
coupled system of them. Since this seems a hopeless goal usually only
one--particle systems have been examined in the form--factor bootstrap
approach.

It is more efficient to examine only the form--factors corresponding
to one particle, and consider all possible constraints on them. That
is one needs to consider all possible fusion processes which return to
particle 1. At a first look it seems that no simplification of the
problem has been obtained. Though we will see in this section that for
all models $\M_{2,2p+3}$ only {\em one} additional constraint equation
for the form-factors involving particle 1 appear.

Let us first examine the fusion rules of this model. They are given by
\cite{Chicago}
\bea  u_{ab}^{|a-b|} = (1-|a-b| \frac\alpha 2)  \sb &,&
\sa u_{ab}^{\min(a+b,2p+1-a-b)} = (a+b) \frac\alpha 2 \sb ,\\
\bar{u}_{ab}^{|a-b|} = |a-b| \frac\alpha 2  \sb &,&
\sa \bar{u}_{ab}^{\min(a+b,2p+1-a-b)} =1- (a+b) \frac\alpha 2 \sb .\eea

We have to find out which kind of fusion processes lead to independent
recursion relations.  In order to understand these techniques consider
the simple fusion process
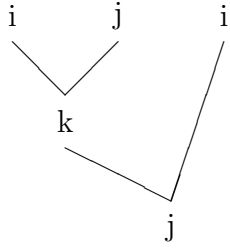
\begin{center}
\begin{picture}(150,110)
\put(40,100){\makebox(0,0){i}}
\put(80,100){\makebox(0,0){j}}
\put(120,100){\makebox(0,0){i}}
\put(40,90){\line(1,-1){20}}
\put(80,90){\line(-1,-1){20}}
\put(60,60){\makebox(0,0){k}}
\put(60,50){\line(2,-1){40}}
\put(120,90){\line(-1,-3){20}}
\put(100,20){\makebox(0,0){j}}
\end{picture}
\end{center}
which is a generalization of the first process of fig. \ref{fig-exam}.
It corresponds to applying twice the bound state equation
(\ref{bounds}).  Analyzing the rapidity shifts, one finds that the
corresponding recursion relation reads as $$ (-1) \lim_{\tilde{\beta}
\to \beta} \lim_{\hat{\beta} \to \beta} (\tilde{\beta}
-\beta)(\hat{\beta}-\beta) $$ $$ F_{n+2} (\hat{\beta}+i \bar
{u}_{i,k}^j+i \bar {u}_{i,j}^{k},\,
\tilde \beta-i \bar {u}_{j,k}^i+i \bar {u}_{i,j}^{k},\,
\beta-i \bar{u}_{j,k}^i,\,
\beta_1,\dots,\beta_{n-1}) =
$$
\be
=\Gamma_{i,j}^{k} \Gamma_{k,i}^j F_n(\beta,\beta_1,\dots,\beta_{n-1})
\sb .\ee
Because of the relation $$ \bar {u}_{i,k}^j+\bar {u}_{i,j}^k+\bar
{u}_{j,k}^i=\pi\sb , $$ we see that as in the example of the
$\M_{2,7}$ model a shift of $i \pi$ between the first and third
rapidity--values appears and therefore the corresponding recursion
relation will reduce to the kinematical one (\ref{rec}).

In order to realize the above process in the perturbed $\M_{2, 2p+3}$
models it is clear that we need to identify the particle $k$ as the
particle $|i-j|$. This means we have considered a specific process
where we have {\em lowered} the particle number. It is therefore useful
to introduce the notation of a {\em minimal fusion process}. Such a
process is defined as one which starts from a set of initial particles
$a_1,a_2,\dots a_k$ and leads to the particle $\sum_{i=1}^k a_i$ using
{\em only} the fusion rules $a_i \times a_j \to a_i+a_j$. Such minimal
fusion processes have the property that independent of how the
particles are fused, the rapidity shifts entering the corresponding
bound state recursion relation are the same. This is seen by analyzing
the rapidity shifts in (\ref{bounds}). In a minimal fusion process
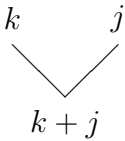
\begin{center}
\begin{picture}(150,60)(0,60)
\put(40,100){\makebox(0,0){$k$}}
\put(80,100){\makebox(0,0){$j$}}
\put(40,90){\line(1,-1){20}}
\put(80,90){\line(-1,-1){20}}
\put(60,60){\makebox(0,0){$k+j$}}
\end{picture}
\end{center}
the rapidities get shifted as $\beta_1 \to \beta_1+ j \frac {i\alpha}
2 $ and $\beta_2 \to \beta_2 -k \frac {i\alpha} 2 $. Since the minimal
fusions are additive the first rapidity in fusioning $a_1,a_2,\dots
a_k$ will have the shift $\beta_1 \to \beta_1 + (a_2+a_3 +\dots +a_k)
\frac{i\alpha} 2$, or in general the shift will be
\be\beta_j \to \beta_j +(-a_1-a_2 \dots -a_{j-1}+a_{j+1} +\dots
+ a_k)\frac{i\alpha} 2\sb .\label{minfus}\ee

The importance of the concept of the minimal fusion processes lies in
the fact, that for the models under consideration one can exclude all
non--minimal processes. That is, one can show that for any
non--minimal fusion process the set of rapidity values contains a
subset which is equal to the minimal one.

To prove this statement assume we have in our fusion process a
subprocess of the type
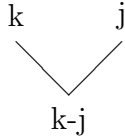
\begin{center}
\begin{picture}(80,50)(0,60)
\put(40,100){\makebox(0,0){k}}
\put(80,100){\makebox(0,0){j}}
\put(40,90){\line(1,-1){20}}
\put(80,90){\line(-1,-1){20}}
\put(60,60){\makebox(0,0){k-j}}
\end{picture}
\end{center}
where we have for simplicity taken $k>j$.  Further assume that we have
formed these particles out of particles $1$ in a minimal way, {\em
i.e.} the subprocess considered is the first in our fusion-graph of a
non--minimal kind.  Then, by (\ref{minfus}) we know the rapidity
shifts which are involved forming particles $k$ and $j$. They are
$$\beta_1\to \beta_1 +(k-1)\frac{i\alpha} 2,\sa\beta_2 \to
\beta_2+(k-3) \frac{i\alpha} 2\sa\dots\sa
\beta_k\to\beta_k-(k-1)\frac{i\alpha} 2$$
and similar for the particle $j$. Now through the final fusion $k
\times j \to k-j$ the first set of rapidities gets further shifted by
$j \frac{i\alpha} 2$ while the second set undergoes a shift of $-i\pi +
k \frac{i\alpha} 2$. This fusion process with its rapidities is shown
in fig. \ref{fig-two}, from which one can easily see that the final
set of rapidity-shifts contains those ones which one obtains if one
forms the particle $k-j$ in a minimal way. Further it is interesting
to note that the excessive shifts pair up in rapidity differences of $
i \pi$ and therefore all these fusion processes reduce to the
kinematical recursion relation.

\footnotesize
\begin{figure}[hbt]
\begin{picture}(450,80)
\put(0,40){\line(0,1){7}}
\put(220,40){\line(0,1){7}}
\put(0,40){\line(1,0){220}}
\put(0,50){$\beta_1 + (k-1) \frac{i\alpha} 2$}
\put(100,50){$\dots$}
\put(130,50){$\beta_k - (k-1) \frac{i\alpha} 2$}
\put(100,25){$ +\, j \frac{i\alpha} 2$}
\put(0,0){$\beta_1 + (k+j-1) \frac{i\alpha} 2$}
\put(100,0){$\dots$}
\put(130,0){$\beta_k+ (j-k+1) \frac{i\alpha} 2$}
\put(230,40){\line(0,1){7}}
\put(450,40){\line(0,1){7}}
\put(230,40){\line(1,0){220}}
\put(230,50){$\beta_{k+1} + (j-1) \frac{i\alpha} 2$}
\put(330,50){$\dots$}
\put(360,50){$\beta_{k+j} - (j-1) \frac{i\alpha} 2$}
\put(310,25){$ -\,(\pi- k \frac{i\alpha} 2)$}
\put(216,0){$\beta_{k+1} -i\pi+ (k+j-1) \frac{i\alpha} 2$}
\put(340,0){$\dots$}
\put(360,0){$\beta_{k+j} -i\pi+ (k-j+1) \frac{i\alpha} 2$}
\end{picture}
\caption{Rapidity shifts for non-minimal fusion process}
\label{fig-two}
\end{figure}

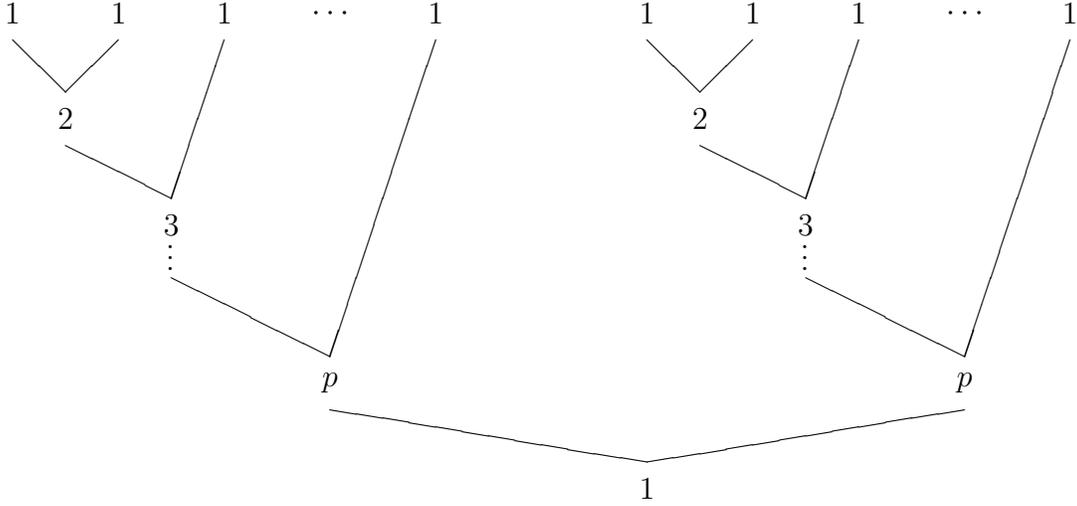
\begin{figure}
\begin{picture}(400,200)(0,-80)
\put(0,0){.}
\put(-40,0){\begin{picture}(150,100)(-40,0)
\put(40,100){\makebox(0,0){1}}
\put(80,100){\makebox(0,0){1}}
\put(120,100){\makebox(0,0){1}}
\put(160,100){\makebox(0,0){$\cdots$}}
\put(200,100){\makebox(0,0){1}}
\put(40,90){\line(1,-1){20}}
\put(80,90){\line(-1,-1){20}}
\put(60,60){\makebox(0,0){2}}
\put(60,50){\line(2,-1){40}}
\put(120,90){\line(-1,-3){20}}
\put(100,20){\makebox(0,0){3}}
\put(98,3){$\vdots$}
\put(200,90){\line(-1,-3){40}}
\put(160,-30){\line(-2,1){60}}
\put(160,-40){\makebox(0,0){$p$}}
\end{picture}
}
\put(200,0){\begin{picture}(150,100)(-40,0)
\put(40,100){\makebox(0,0){1}}
\put(80,100){\makebox(0,0){1}}
\put(120,100){\makebox(0,0){1}}
\put(160,100){\makebox(0,0){$\cdots$}}
\put(200,100){\makebox(0,0){1}}
\put(40,90){\line(1,-1){20}}
\put(80,90){\line(-1,-1){20}}
\put(60,60){\makebox(0,0){2}}
\put(60,50){\line(2,-1){40}}
\put(120,90){\line(-1,-3){20}}
\put(100,20){\makebox(0,0){3}}
\put(98,3){$\vdots$}
\put(200,90){\line(-1,-3){40}}
\put(160,-30){\line(-2,1){60}}
\put(160,-40){\makebox(0,0){$p$}}
\end{picture}
}
\put(160,-50){\line(6,-1){120}}
\put(400,-50){\line(-6,-1){120}}
\put(280,-80){\makebox(0,0){1}}
\end{picture}
\caption{Fusion--graph for the bound state recursive equation}
\label{fig-three}
\end{figure}

\normalsize
We have analyzed of how the rapidity-shifts behave if we form a
particle $k$ out of particles 1. Our goal though is to return to
particle 1. For this scope there are two possible processes, namely
$a_k \times a_{k+1} \to 1$ and $a_p \times a_p \to 1$. The first one
can be excluded as a special case of the above consideration, since it
reduces to the kinematical recursion relation. Therefore remains only
one unique way to return to particle 1. But we have shown above that
forming the particle $p$ it is sufficient to consider minimal fusion
processes and further that the rapidity shifts of that process are
unique. Therefore we can conclude that there is exactly {\em one}
recursion-relation (besides the kinematical one) giving constraints
for the form--factors of the particle one.  The corresponding
fusion-graph is shown in fig. \ref{fig-three}.

It involves $2p$ rapidity shifts which take the values
\be \hat{\beta}_1+i(2p-1)\frac{i\alpha} 2, \sa
\hat{\beta}_2+i(2p-3)\frac{i\alpha} 2, \sa \dots ,
\hat{\beta}_{2p-1}-i(2p-3)\frac{i\alpha} 2, \sa
\hat{\beta}_{2p}-i(2p-1)\frac{i\alpha} 2\ee

The final recursion relation reads as $$(-i)^{(2p-1)}
\lim_{\stackrel{\hat{\beta}_i\to\hat{\beta}_{i+1}}{i=1,\dots,p-1}}
\,\lim_{\stackrel{\hat{\beta}_i\to\hat{\beta}_{i+1}}{i=p+1,\dots,2p-1}}
\lim_{\hat{\beta}_p\to\hat{\beta}_{2p}}
\,(\hat{\beta}_{p} -\hat{\beta}_{2p})
\prod_{i=1}^{p-1} (\hat{\beta}_{i}-\hat{\beta}_{i+1})
\prod_{i=p+1}^{2p-1} (\hat{\beta}_{i}-\hat{\beta}_{i+1})\times
$$ $$ F_{n+2p-1} \left (\hat{\beta}_{1}+(2p-1)\frac {i\alpha} 2,
\,\hat{\beta}_{2}+(2p-3)\frac {i\alpha} 2,
\dots,
\hat{\beta}_{2p}-(2p-1)\frac {i\alpha} 2,
\,\beta_1,\dots,\beta_{n-1}\right ) =$$
\be=\prod_{k=1}^{p-1} (\Gamma_{1k}^{k+1})^2 \, \Gamma_{pp}^1
\,\,F_n (\hat{\beta}_{2p},\beta_1, \dots,\beta_{n-1})
\sb .\label{rec-gen}\ee

For consistency we have to check whether the product of on--shell
vertices $\Gamma_{ab}^{c}$ on the right hand side gives the same
result for any minimal fusion process. Note that since minimal fusions
map always $a\times b \to a+b$ the on-shell vertex depends effectively
only on two variables $\Gamma_{ab}$. Using the expression for the
$S$--matrix (\ref{sss}) and the definition for the on--shell vertices
(\ref{residueS}) we find (for simplicity of notation we put $a>b$)
\be
\left ( \Gamma_{ab} \right )^2 = 2 \tan (a+b) \frac \pi{2p+1} \,
\frac{\tan a\frac\pi{2p+1}}{\tan b\frac \pi{2p+1}} \prod_{k=1}^{b-1}
\left ( \frac{\tan(a+k)\frac \pi{2p+1}}{\tan (b-k)\frac \pi{2p+1}}
\right )^2 \sb .\ee
Using this expression one can show that the algebra formed by the
on--shell vertices is associative, {\em i.e.}
\be \Gamma_{a_1,a_2} \Gamma_{a_1+a_2,a_3} = \Gamma_{a_1,a_2+a_3}
\Gamma_{a_2,a_3} \sb ,\ee
and therefore we conclude that any minimal fusion--path gives the same
constant on the right-hand side of equation (\ref{rec-gen}).

We have therefore shown that all possible constraints arising from
fusion processes can be encoded in exactly {\em one} recursive
equation for the form--factors of the particle 1. Note that this is a
considerable simplification in confrontation with the coupled recusive
equations for form--factors with indices corresponding to different
particles with which we started with.

\subsection{Polynomial recursive equation and operator content of
perturbed $\M_{2,2p+3}$ models}

In section \ref{sec-SHG} we have classified the solutions of the
kinematical recursion relation (\ref{rec}). We found a tower-like
structure growing with the number of particles in the form-factor.  In
the examples of Yang--Lee and $\M_{2,7}$ models the dimensionality was
severely restricted by the bound state equations, which we imposed in
addition to the kinematical recursion relation. We want to carry out
this analysis now for the theories $\M_{2,2p+3}$ in general. As a
first step we need to calculate the recursion relation for $Q_n$
corresponding to (\ref{rec-gen}). In terms of the the parametrization
(\ref{para}) it is a tedious but nevertheless straightforward
calculation. The final result is given by $$ Q_{n+2p-1}
(x\omega^{\frac{2p-1}2},x\omega^{\frac{2p-3}2},\dots,
x\omega^{\frac{1}2},x\omega^{-\frac{1}2},\dots,x\omega^{-\frac{2p-1}2},x_1,
\dots,x_{n-1}) =
 $$
\be
=U_n^p (x,x_1,\dots,x_{n-1}) Q_n(x,x_1,\dots,x_{n-1})
\sb ,\ee
where we have defined $x \equiv e^{\hat{\beta}_{2p}}$. The function
$U_n^p$ is given by $$ U_n^p(x,x_1,\dots,x_{n-1}) = (-1)^{\frac{p
(p+1)}{2}+1} \prod_{k=2}^{p-1} [k]^2 \times $$
\be \times
 x^{p(2p-1)}
\prod_{i=1}^{n-1} (x+x_i)
\prod_{k=2}^p \prod_{i=1}^{n-1} (x-x_i \omega^k)(x-x_i
\omega^{-k}) \sb .\label{BS}\ee

We want to determine the operator content of the theory. Note that the
recursion relations (\ref{BS}) link the even und odd sectors of the
form--factors. In fact they relate the form--factors $F_{n+2p-1}$ to
$F_{n}$. We know from section \ref{sec-SHG} that without considering
the bound--state axiom of the form--factors the solution space grows
linearly which is a consequence of the kinematical recursion relation.

We want to determine the subspace which in addition fulfills the
relation (\ref{BS}). With the recursion relation (\ref{BS}) at hand
this is now a rather easy task. The argument for the truncation of the
solution space carries over straight forward from our examples.  The
recursion relation (\ref{BS}) acts only on polynomials $Q_n$ with $n
\geq 2p$. Therefore up to this level we have to consider only the
kinematical recursion relation. At level $2p$ the even and odd sectors
are linked which are independent by (\ref{rec}).  Only for $Q_n$ with
$n>2p$ the bound state relation constrains the dimensionality of the
solution space. Now consider $Q_{n},\, n>2p$. Its dimensionality is
determined by the kernel of the recursion relations, but now of both,
{\em i.e.} (\ref{rec}) and (\ref{BS}). It is given by
\be
\K_n = \prod_{i<j} (x_i+x_j) (x+\omega x_i)(x+\omega^{-1} x_i)
\sb .\ee
This function though is not of the degree $n(n-1)/2$ as required from
Lorentz invariance. This means the kernel is zero-dimensional, and no
additional solutions for the recursion relations exist at any level
$n>2p$. Therefore the dimensionality of the solution space is
determined by that of $Q_{2p}$ which is $p$.  Similary the argument we
have applied for $\M_{2,7}$ equally carries over for all theories
$\M_{2,2p+3}$.

These arguments show that there exist only $p$ independent scalar
`primary` operators in the perturbed minimal models $\M_{2,2p+3}$.
This is the same amount as there are in the conformally invariant
theories.  A base in this space is given by the elementary solutions
(\ref{element}) $Q_n (k)$ with $k=1,\dots,p$.  We denote the operators
corresponding to these solutions as $\Psi_k$.

\subsection{Physical Interpretation \label{sec-six}}

We conjecture that the set of solutions $\Psi_k$ correspond to the
off-critical fields which in the ultraviolet limit turn into the
primary fields $\phi_{1,2 k +1}$.

There are several observations which support this hypothesis.  The
polynomial $Q(1)$ factors as $$Q(1)(x_1,\dots x_n)= \s_1 \s_{n-1}
\times Q'(x_1,\dots,x_n)\sb ,$$ $Q'$ being a polynomial,
 which is seen by analyzing the expression of the determinant
(\ref{element}).
Using the fact that $$\sum_j x_j^{-1} =
\frac{\s_{n-1}}{\s_n} \sb ,$$ and the sum rule $$\sum_{k=1}^n \cos (2k-1) x
= \frac 12 \frac{\sin (2nx)}{\sin x}\sb , $$
this implies that in general the form-factors
corresponding to these solutions will factor as $$F_{\e_1,\dots\e_n} =
\sum (m_{\e_j} x_j) \sum (m_{\e_j} x_j^{-1})
\times F'_{\e_1,\dots\e_n} \sb . $$
This is the requirement
(together with apropriate normalization) for the form-factors to be
interpreted as corresponding to the trace of the energy momentum
tensor $\Theta =\frac 14 T^\mu_\mu$. But since in the perturbed models
\cite{Zam} $$ \Theta =
\lambda (\Delta -1) \phi_{1,3}\sb ,$$ we can identify these form-factors as
belonging to the operator $\phi_{1,3}$ up to some normalization.
Further, Smirnov has identified two exponential operators in
\cite{smirnov13}, namely $Pe^{\frac i2\sqrt{\gamma}\phi}P$ and
$Pe^{i\sqrt{\gamma}\phi}P$. Even if we were not able to rewrite his
integral representation in terms of our determinant form we have
checked for several values of $p$ and $n$ that his form-factors are
equivalent to ours for $\Psi_1$ and $\Psi_p$. In the scaling limit
these two operators turn into the conformal fields $\phi_{1,3}$ and
$\phi_{1,2}$ respectively.

The perturbed minimal models $\M_{2,2p+3}+\phi_{1,3}$ correspond to
the reduced Sine-Gordon model at the rational values of the coupling
constant $\gamma$. Even if we did not use this correspondence in the
derivation of the form-factors, it is useful to use this property in
the physical interpretation of the operators $\Psi_k$. The operators
in the reduced models correspond to those in Sine-Gordon theory when
projected onto the soliton-free sector. That is, the fields
$\phi_{1,k}$ can be identified with the operators $P e^{i \frac k2
\sqrt{\gamma} \phi} P$, where $P$ denotes the projection and $\phi$ is
the Sine-Gordon field \cite{smirnov13}. Therefore for a consistent
interpretation of our solutions we expect $$Q_n (k) = Q_n
(p-\frac{2k-1}2)\sb .$$ That this is indeed the case follows directly
from the symmetry properties of the symbols $[m]$ appearing in the
expressions for the determinant $Q(k)$.

As a final point let us investigate the cluster property of the
operators $\Psi_k$.  By cluster transformation we generally mean the
behaviour of a form factor under the shift of a subset of the
rapidities, {\em i.e.}
\EQ
F_n^{\O_a}(\beta_1+\Delta,\dots,\beta_m+\Delta,\beta_{m+1},\dots,
\beta_{n}) \,\,\, .
\label{cluss}
\EN
Taking the limit $\Lambda\rightarrow\infty$, $F_n^{{\cal O}_a}$ can be
decomposed into two functions of $m$ and $(n-m)$ variables
respectively, where both functions satisfy all the set of axioms for
the form-factors.  Therefore they can be considered as FF of some
operators ${\cal O}_b$ and ${\cal O}_c$
\be
\lim_{\Lambda\rightarrow \infty}
F_n^{\O_a}(\beta_1+\Delta,\dots,\beta_m+\Delta,\beta_{m+1},\dots,
\beta_{n})
= F_m^{\O_b}(\beta_1,\dots,\beta_m)
F_{n-m}^{\O_c}(\beta_{m+1},\dots,\beta_n)
\ee
We denote this operation as
\[
{\cal O}_a\, \rightarrow \,{\cal O}_b \times {\cal O}_c \,\,\, .
\]
We will prove that the operators $\Psi_k$ are mapped onto themselves
under the cluster transformation, {\em i.e.}
\EQ
\Psi_k \,\rightarrow \, \Psi_k \times \Psi_k \,\,\, .
\EN
We define the cluster-operator $\C_m$ (acting on the symmetric
functions) by means of
\be
\C_m \left ( f(x_1,\dots,x_n) \right ) \equiv f(x_1e^\D,x_2e^\D,\dots,
x_me^\D,x_{m+1},\dots,x_n ) \sa m<n \sa .
\ee
Further, using the notation $$
\hat \s_i^{(n-k)} \equiv \s_i^{(n-k)} (x_{n-k+1},x_{n-k+2},\dots,x_n) \sb.
$$ one finds that
\be
\C_m(\s_k^{(n)})= \sum_{i=1}^{k} \s_{k-i}^{(m)} e^{(k-i)\D} \hat \s_i^{(n-m)}
\sb .\label{csigma}
\ee
Since the cluster properties are fixed by the leading term of this
sum, we have
\be
\begin{array}{ll}
\C_m(\s_k^{(n)})\sim \s_m^{(m)} e^{m\D} \hat \s_{k-m}^{(n-m)} & m \leq
k \sb ,\\
\C_m(\s_k^{(n)})\sim \s_k^{(m)} e^{k\D} & m \geq k \sb .
\end{array}
\label{clussym}
\ee
Now let us consider separately the cluster property of each term
entering their parametrization
\be
F_n^k(\beta_1,\dots,\beta_n) = H_n^k \,Q_n(k)\,
\prod_{i<j}^n \frac{F^{\rm min}(\beta_{ij})}{(x_i+x_j)
\sinh \frac 12 (\beta_{ij} -i \pi \alpha)
\sinh \frac 12 (\beta_{ij} +i \pi \alpha)}  \label{fn}\,\,\, .
\ee
Since $$F^{\rm min}(\beta) \stackrel{\beta\to\infty}{\goto} -
\frac 1{4 \zeta (i\pi) \pi} e^{\beta} $$
 we have
\be
\prod _{i<j}^n F^{\rm min}(\beta_{ij}) \goto
\prod _{i<j}^m F^{\rm min}(\beta_{ij})
\prod _{i<j=m+1}^n F^{\rm min}(\beta_{ij})
(\frac {-1}{4 \zeta (i\pi) \pi})^{m(n-m)} \prod_{i=1}^{m}
\prod_{j=1}^{n-m} e^{\beta_{ij}+\Delta} \,\,\,.
\label{first}
\ee
Further, using eq.\,(\ref{clussym}), the cluster property of the
elementary solution $Q_n(k)$ is given by \cite{a10}
\be
\C_m(\frac{Q_{n}(k)}{\prod_{i<j}^n (x_i+x_j)} ) \sim  [k]
\frac{Q_{m}(k)}{\prod_{i<j}^m (x_i+x_j)}\frac{Q_{n-m}(k)}{\prod_{i<j}
^{n-m} (x_i+x_j)}
\sb ,\label{cq}
\ee
Finally decomposing the factor $$\prod_{i<j}^n \sinh \frac 12
(\beta_{ij} -i \pi \alpha)
\sinh \frac 12 (\beta_{ij} +i \pi \alpha) \sb , $$
and using the values of the constants $H_n$ (\ref{hn}) we find $$
\C_m(F_n^k(\beta_1,\dots\beta_n)  = \frac{[k]}{C_0}
F_m^k(\beta_1,\dots\beta_m) F_{n-m}^k(\beta_1,\dots\beta_{n-m}) $$ we
conclude that the FF of $\Psi_k$ are mapped onto themselves under the
cluster transformation if $C_0 = [k]$.  Since this is a distinguished
property of exponential operators \cite{smirnov13,nankai},
it is natural to identify the operators $\Psi_k$ with the fields
$Pe^{k \sqrt{\gamma} \phi}P$.

\resection{Descendent Operators \label{sec-desc}}

A crucial point in our analysis was the strict interpretation of the
form-factor axioms leading to the ansatz (\ref{ansatz}) for $Q_n$. In
order to check the consistency one should also examine descendent
operators in the theory. We will here only sketch the problem, a
detailed account will be given elsewhere \cite{descend}.

For simplicity we discuss the Yang-Lee model. We use the notation that
$\bar{x} = 1/x$ and $\bar{\s}_i =\s_i(\bar{x}_1,\dots\bar{x}_n)$.
Simple derivative operators can be obtained by  multiplying $Q_n$ by
$\s_1^m$ and $\bar{\s}_1^m$. Because of the identity
$$\bar{\s}_i(x_1,\dots,x_n) = \frac{\s_{n-i}(x_1,\dots,x_n)}{\s_n
(x_1,\dots,x_n)}\sb ,$$
any $\bar{x}$ dependence can be rewritten in terms
of $x$ , retaining the form of $Q$ as in ansatz (\ref{ansatz}). We
will therefore restrict this discussion to only `chiral'
descendents which we want to compare with the structure of the Verma
module created by $L_{-n}$ leaving out the $\bar{L}_{-n}$dependence.

As discussed in section \ref{sec-para} we cannot expect that ansatz
(\ref{ansatz}) gives a direct correspondance beetween conformal and
descendent states. Nevertheless let us examine the operator content
of polynomial solutions for $Q_n$ for several spin values $s$.
The strategy is the same as for the perturbed primary operators. We
solve the relations (\ref{rec}) and (\ref{bsyl}) recursively using the property
that the space of solutions at level $n$ is given by the number of
solutions at level $n-1$ plus the dimension of the kernel of the
combined recursion relations. In table \ref{tab-1} we have written
down the degree for the Polynomials $Q_n$ required from Lorentz
invariance in comparison with the dimensions of the kernel of the
recursion relations.
\begin{table}
\begin{center}\begin{tabular}{||c|c|c||} \hline $ n$&
$deg (Q_n^{(s)})=\frac 12 n(n-1)+s$ & $deg \K = \left \{ \begin{array}{c}
\frac 12 n(n-1)\sa n=1\\ \frac 32 n(n-1) n\geq 2 \end{array} \right .$
\\
\hline
$1$ &$ 0+s$ &$ 0$ \\
$2$ &$1+ s$&$3  $\\
$3$ &$3 +s$&$ 9 $\\
$4$ &$6 +s$&$ 18$ \\
$5$ &$10+s$ &$30$  \\
$6$ &$15+s$ &$45$  \\
$7$ &$21+s$ &$63$  \\
\hline\end{tabular}\end{center}
\caption{Comparing the total degrees of the polynomials $Q_n$ and the
Kernel of the recursion relations for spin $s$ and level $n$}
\label{tab-1}
\end{table}
 Let us now count the independent solutions for various spin levels.
For $s=1$ the only solution is that generated by $Q_1= A_1\s_1$, which
corresponds exactly to the level 1 descendent of the field $\phi$ of
the Yang-Lee model. For spin 2 we have two possible solutions,
generated by the initial polynomials
$$ Q_1= A_1\s_1^2 \sb Q_2 =A_2\K_2 \sb .$$ Similar higher spin values can
be investigated. Confronting the dimensions in table \ref{tab-1} we
see that the first time a Kernel solution for $n=3$ occurs for spin
$s=6$. Let us examine the solution space for that particluar spin
value. The solutions are
\bea Q_1 &=& A_1\s_1^6 \nonumber \\
Q_2 &=& (A_2 \s_1^4 + A_3 \s_2 \s_1^2 +A_4 \s_2^2) \K_2 \nonumber\\
Q_3 &=& A_5 \K_3 \eea
that is we have 5 independent operators.

Now let us confront this situation with conformal field theory.
The number of descendent operators
is just given by the character expansions,
\bea
\chi_{1,1} &=& 1+q^2+q^3+q^4+2 q^5 +2 q^6 +O(q^7) \nonumber\\
\chi_{1,3} &=& 1+q+q^2+q^3+2 q^4+2 q^5 +3 q^6 +O(q^7) \sb .\eea
Summing up the values we have in CFT 1 descendent operator at spin 1,
2 operators at spin 2 and 5 operators at spin 6. We confront these
values with the ones in the perturbed model and find they coincide.

We have presented here this counting argument just for a few spin
values. Using the Rogers-Ramanujan identities,
 one can prove \cite{descend} that the result holds in general.
The same proceedure can be adopted for the other
$\phi_{13}$ perturbed models discussed in this paper. There the
counting becomes more difficult since the bound state relation
connects $Q_{n+2p}$ with $Q_{n}$ and therefore leaves the possibility
of intermediate steps of the kinematical recursion relation giving an
extra freedom of parameters. We do not yet have general results for
these models, but in the cases we have investigated, the
dimensions of the spaces of descendent operators in the perturbed and
the conformal models coincide.

\resection{Conclusions \label{sec-conc}}

We have constructed and analyzed the form-factors for the class of
models $M_{2,2p+3} +\phi_{1,3}$. We chose these models since they are
the simplest multi-particle systems containing only scalar states.  We
constructed the full set of form-factors for `primary operators' and
found that the operator content of these perturbed models is
isomorphic to that of the conformal ones. The proof uses the fact that
the form--factor equations do not refer to a specific operator
and therefore the classification of solutions gives a means of
determining the operator content of the theory.

In this analysis it is interesting to note the role of the various
equations. The Watson's equations (\ref{wat1}) and (\ref{wat2}) can be
solved by conveniently parametrizing the form-factors. The
kinematical recursion equation (\ref{kin}) defines a constraint for
the form-factors and its solutions have a tower like structure, but
still admit an infinite number of `primary fields', which we
defined to be fields with the lowest scaling dimensions in the
ultraviolet limit.  As long as we do not enforce the bound state
equation (\ref{bounds}) the solution space of the form-factor
equations is identical to that of the Sinh--Gordon model. The
reduction of the solution space to finite dimensionality is achieved
by the bound state axiom which in this phenomenological approach takes
the role of the quantum group reduction mechanism. It would be
interesting to understand whether this is a generic feature of the
form-factor equations for perturbed minimal models.

We found a base in this space of solutions which we conjecture to
correspond to the off-critical extension of the conformal primary
fields $\phi_{1,2k+1}$. We have given several arguments in support of
this identification. Since the form-factor expansion corresponds to a
large distance expansion, the ultraviolet limit is not very tractable,
since the whole series should be resummed. We also carried out several
numerical calculations in ordre to determine the scaling dimensions of
the operators $\Psi_{k}$ but did not obtain conclusive answers,
because of the complexity of the functional dependence of the
form-factors on the rapidity variables.  A final verification of this
identification should be done in an algebraic framework.

\subsection*{Acknowledgments}
I would like to thank A. Fring and M. Niedermaier for helpful comments
and N. Mackay for a critical reading of the manuscript. This work was
supported by SERC grant GR/J20661.

\end{document}